\documentclass[
      preprint,
floatfix,
amssymb,
amsmath,
aps,
pra,showpacs]{revtex4}

\usepackage{times}
\usepackage{graphicx}

\newcommand{\uwa}{School of Physics, The University of Western Australia\protect\\ Nedlands, Western Australia 6907, Australia}

\newcommand{\reference}[1]{
	\bibitem
	{#1}
}

\newcommand{\Iav}{\ensuremath{I_\mathrm{av}}}
\newcommand{\Ep}{\ensuremath{E_\mathrm{pulse}}}
\newcommand{\recip}[1]{\ensuremath{\frac{1}{#1}}}

\newcommand{\dd}{\;\mathrm{d}}
\newcommand{\Lint}[1]{\int_{-L/2}^{L/2}#1\dd z}
\newcommand{\micron}{\ensuremath{\mu\mathrm{m}}}
\newcommand{\lam}{\lambda}

\newcommand{\apx}{\ensuremath{\approx}}
\newcommand{\Pd}{\ensuremath{P_{\mathrm{det}}}}
\newcommand{\Pav}{\ensuremath{P_{\mathrm{av}}}}
\newcommand{\tec}{technique}
\newcommand{\vb}{vacuum birefringence}
\newcommand{\vp}{vacuum polarization}
\newcommand{\ti}{\ensuremath{\tau_\mathrm{int}}}
\newcommand{\npar}{\ensuremath{n_{\|}}}
\newcommand{\nperp}{\ensuremath{n_{\perp}}}
\newcommand{\fpar}{\ensuremath{\nu_{\|}}}
\newcommand{\fperp}{\ensuremath{\nu_{\perp}}}
\newcommand{\lpar}{\ensuremath{l_{\|}}}
\newcommand{\lperp}{\ensuremath{l_{\perp}}}
\newcommand{\unit}[1]{\ensuremath{\,\mathrm{#1}}}
\newcommand{\ee}[1]{\ensuremath{\times 10^{ #1}}}
\newcommand{\brac}[1]{\left( #1 \right)}

\newcommand{\eq}[2]{\begin{eqnarray}#2\label{#1}\end{eqnarray}}
\newcommand{\makefig}[3]{
	\begin{figure}[bt]
	\centering
	\includegraphics[width=3.4in]{#1}
	\caption{#3}
	\label{#2}
	\end{figure}
}

\begin{document}
\bibliographystyle{apsrev}
\title{Ultrafast Resonant Polarization Interferometry:  Towards the First  Direct Detection of Vacuum Polarization}
\author{Andre N.  Luiten}
\email{andre@physics.uwa.edu.au}
\author{Jesse C. Petersen}
\altaffiliation{Now with Department of Physics, Simon Fraser University, Burnaby, British Columbia, Canada.}
\affiliation{\uwa}
\date{\today}
\begin{abstract}
Vacuum polarization, an effect predicted nearly 70 years ago, is still yet to be directly  detected despite significant experimental effort.  Previous attempts  have  made use of large  liquid-helium cooled electromagnets which inadvertently generate  spurious signals that mask the desired signal. We present a novel approach for the ultra-sensitive detection of optical birefringence that can be usefully applied to a laboratory detection of  vacuum polarization.   The new technique has   a predicted birefringence measurement sensitivity of $\Delta n \sim 10^{-20}$  
in a 1 second measurement. When combined with the extreme polarizing fields achievable in this design  we predict that a  \vp\  signal  will be seen in a measurement of just a few days in duration.     
\end{abstract} %
\maketitle

\section{Introduction}

A number of physical effects of current interest manifest themselves as a birefringence appearing in response to the application of strong electromagnetic fields. One example is the small degree of magnetically induced birefringence that arises in the Cotton-Mouton effect~\cite{Muroo,muroo2,cameron}. In this case light traversing  a medium exposed to a strong transverse magnetic field observes a different refractive index for polarization states parallel and perpendicular to the magnetic field direction. In dilute gases the Cotton-Mouton effect can be exceedingly small~\cite{Muroo,muroo2,cameron}.  An analogous, but even smaller field-induced birefringence is predicted to occur in vacuum because of corrections to Maxwell's equations arising under quantum electrodynamics (QED)~ \cite{euler,heisenberg,weisskopf,schwinger}. This correction, originally made over 70 years ago, predicts   a  birefringence  that is only of the order of ${\Delta n \sim 10^{-21}}$ for any realistic  laboratory magnetic field.  The weakness of the vacuum birefringence effect has conspired with an unavoidable generation of large spurious signals in the techniques used to date, to prevent a successful detection of vacuum birefringence in the laboratory.

In this article we propose a novel method for making extremely sensitive birefringence measurements  based on the use of frequency-stabilized mode-locked lasers and low dispersion optical resonators. We  will measure the birefringence induced inside a focussed femtosecond duration pulse of light.  This technique appears to hold  the promise of  state-of-the-art sensitivity while using only room-temperature table-top apparatus that is reliable and relatively inexpensive. The approach circumvents the most important disadvantages of conventional approaches, in particular, it avoids the generation of high level spurious signals that mask the desired birefringence signal.  The necessary equipment is commercially available and we believe that a number of laboratories are  well  positioned to commence research in this direction.  This new approach was only made possible because of the recent remarkable developments in mode-locked laser frequency stabilization techniques~\cite{diddams}.


We commence this article with a short description of existing methods for ultra-sensitive detection of field-induced birefringence and contrast this with the new ultrafast approach.  We then present a comparative analysis of the sensitivity of the new and conventional approaches. Finally we consider the consequences of applying   the new technique to vacuum polarization measurements. 
We demonstrate that not only does the new technique avoid generation of spurious signals, but its sensitivity is comparable to the best previously reported.  In addition, the degree of polarization that is achievable with a resonant short pulse of light is comparable to the highest values achievable with the conventional approach. 



\section{Background}
Traditionally experiments aimed at detecting a weak field-induced birefringence make use of high intensity static or low-frequency oscillating magnetic fields supplied by extremely powerful superconducting electromagnets~\cite{iacopini,cameron, pvlas1,lee,taiwan1,bmv, taiwan2, boer,hall,muroo2,Muroo}. The induced birefringence is observed by sending a linearly polarized field through the magnetic field and observing the modification of its polarization state (ellipsometry). Superconducting  electromagnets can supply extremely intense fields ($5 -  25$\,T) and are thus useful because they create high levels of polarization, but also unfortunately posses a number of key limitations.  The most obvious  of these disadvantages is that they are large and operationally expensive while  the generated fields can only be modulated at low frequencies. This limitation on modulation frequency means that any birefringence signal can easily be buried in the low frequency noise of the detector  necessitating  the use of more elaborate modulation schemes~\cite{cameron, pvlas1,taiwan1}. Of even more consequence  for highly sensitive experiments are the unfortunate results of the   large volume fields generated by the magnets, and the high forces that are intrinsically part of high-energy superconducting magnet systems. The high forces result in movement of the optical elements in  the detection system which can masquerade as a birefringence signal~\cite{cameron,itnoise}. The unconfined nature of the magnetic field makes it difficult to properly shield  the detection apparatus and this is problematic because low levels of residual field can act  on the detection system components so as to generate a false  birefringence signal~\cite{lee,bmv,cameron,itnoise}.  Existing searches for  vacuum birefringence were limited by these types of spurious signals.

On its face an attractive alternative to high energy   magnets would be the use of optical fields to generate the polarization necessary for the experiment.  A number of authors have suggested the use of continuous-wave (cw) lasers to generate the necessary fields, however,  the energy density of these optical fields is extremely small in comparison with that of the superconducting magnet generated fields ~\cite{vpol,alek,partovi}.  In this paper we propose to use extremely intense short pulses of optical radiation to generate the high fields necessary to polarize the media. As has already been noted~\cite{lee}, the peak magnetic fields that exist within these intense short pulses of light (of the order of $10^{5}$\,T for a 1\,J, 50\,fs pulse focused into 10$^{-12}$m$^2$) can greatly exceed the fields that can be generated by any other means.  The high degree of confinement of the optical field means that although the peak electromagnetic fields are very high, the total energy stored in the field is much smaller than a static magnetic field that would produce  an equivalent birefringence signal. The pulsed light technique thus has twin benefits in that it  eliminates any large forces from the experiment, and also  make shielding of the detection apparatus from the strong fields very simple. The obvious disadvantage of this approach is that the high fields only persist for a short period of time in any particular location, and over a very small volume. To overcome this challenge one requires a detection technology with a very high temporal and spatial resolution so as not to average the signal away. In this paper we propose a novel synchronous detection \tec\ that satisfies both of these requirements and which uses highly precise frequency metrology \tec s~\cite{hall}. Our approach will simultaneously resonate the strong field for polarizing the media together with the probing field that detects the resulting birefringence. This has the advantage of allowing simultaneously high intensity fields as well as a high interaction rate.  The combination of a highly sensitive detection \tec\ and high magnitude of  polarization  potentially puts detection of QED vacuum polarization within the grasp of an all-optical tabletop experiment using existing technology.  


\section{Resonant Polarization Interferometry}

J. Hall et al have reported an experimental technique capable of measuring  birefringence  with great precision~\cite{hall}.  We will refer to the device, illustrated schematically in Fig.~\ref{fpbi}, as a Resonant Polarization Interferometer (RPI).  The   technique relies on frequency locking two continuous-wave (cw) but orthogonally-polarized lasers to the same longitudinal mode of a resonator using the Pound-Drever-Hall technique~\cite{pdh,black,hils}.  To first order the fractional frequency difference between the stabilized laser frequencies is equal to the fractional difference in the optical path length of the resonator measured in the two polarization states:

\eq{fracfreqdiff}{\frac{\fperp-\fpar}{\nu_0} = \frac{\lpar-\lperp}{l_0}}
where $\nu_0$ is the average frequency of the two modes and $l_0$ is the average length of the resonator.
 A path length difference will arise from any birefringence in the cavity in addition to that coming from any intrinsic birefringence of the cavity mirror coatings~\cite{hall}:
 
 \eq{fracbirefringence}{\fperp-\fpar  \sim   \frac{ \npar-\nperp}{n_{0}} \nu_0 +\frac{c}{2 n_0 L}  \frac{\delta \phi}{2 \pi}}
where ${c}/({2 n_0 L})$ is the longitudinal mode spacing of the resonant cavity, $\delta \phi$ is the difference in the reflection phase for the two polarisations, and $n_0$ is the average refractive index in  the resonator.  The laser frequency difference, $\fperp-\fpar$, can be extracted by detecting the beat-note between the lasers  and measuring the beat-note frequency with   a conventional high precision frequency counter. 

\makefig{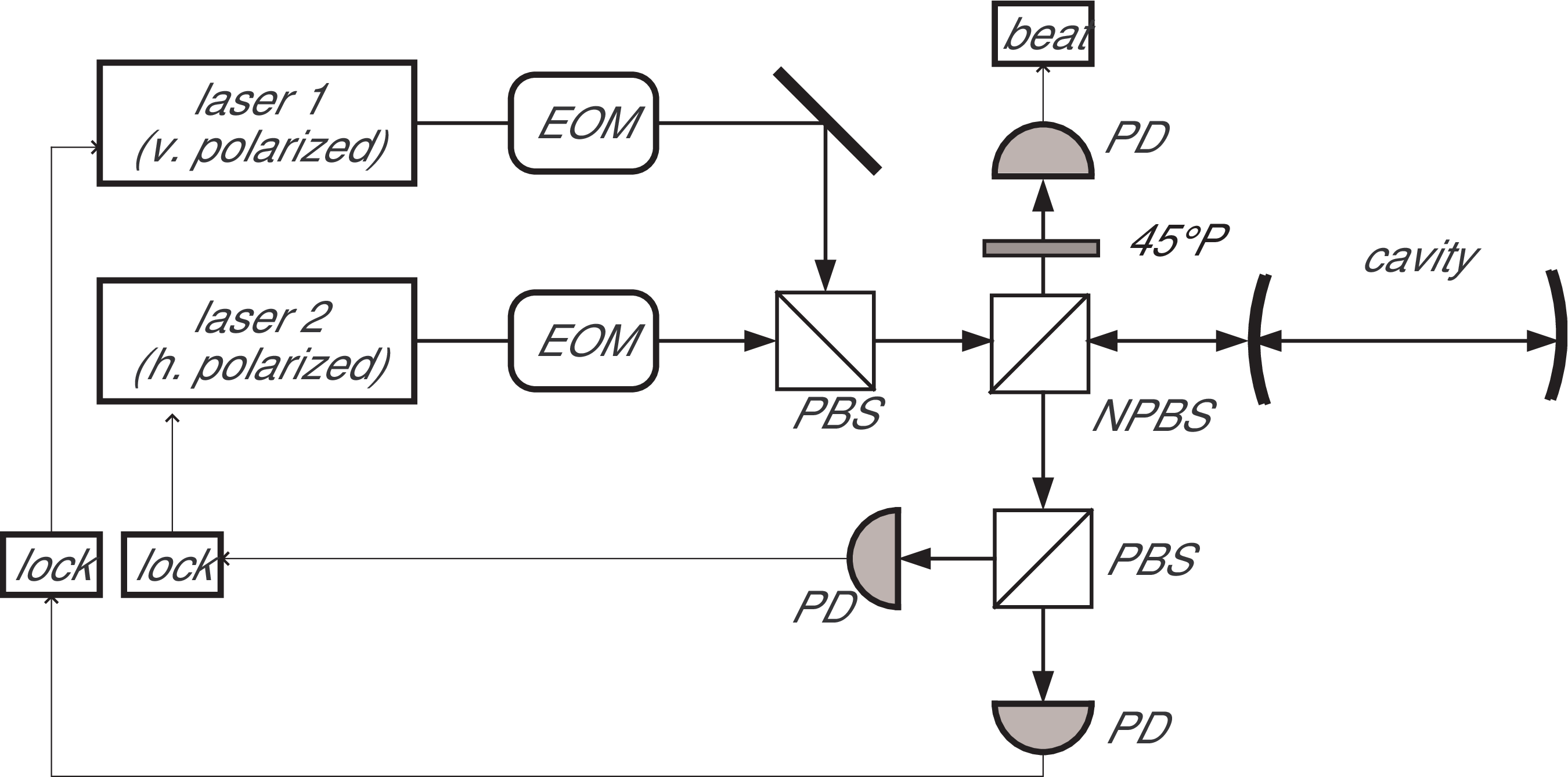}{fpbi}{A Resonant Polarization Interferometer following that described in reference \cite{hall}. PD-photodiode; PBS-polarizing beamsplitter; NPBS-non-polarizing beamsplitter, 
EOM-electro-optic modulator, 45$^{\circ}$P is a polarizer at $45^{\circ}$ to the polarization of lasers}

It is apparent from Eq.~(\ref{fracbirefringence}) that the RPI approach  gives a potentially high sensitivity  since a small fractional difference in the refractive index is multiplied by the optical frequency $\nu_0$ ($\sim 3$~x~$10^{14}$ Hz). In addition, we note that cavity length fluctuations   arising from vibration or temperature fluctuations will be common to both polarizations and hence do not appear in the measured frequency difference signal.  This avoids the need for high quality vibration isolation or temperature control of the detection  resonator. 
 
If technical noise sources such as laser pointing instability and power fluctuations can be adequately reduced, then the key residual fluctuations in the frequency difference signal will be due to the inherent noise in the frequency locking system.  With sufficient servo gain and high modulation frequencies, the dominant residual noise source is photon shot noise. An order of magnitude estimate shows that this will limit the accuracy of each locked laser frequency to a fraction of the resonance bandwidth equal to ~\cite{sal,hall}:
\eq{dshot}{\delta_{\mathrm{shot}} \sim \sqrt{\frac{h\nu}{\Pd \ti\ }}, } 
where $h$ is Planck's constant, $\nu$ is the laser frequency, \Pd\ is the power falling on the feedback photodiode and $\ti$ is the integration time. For a more detailed noise analysis see Sect.~\ref{sec4} below, but as an initial estimate assume the use of   800\,nm laser light and photodiodes that accept a few milliwatts of  incident light. In this case, the laser will be locked to one part in $10^{8}$ of the  cavity bandwidth after 1 second of integration time.  In an optical resonator of length $L$ with finesse $F$, the frequency  bandwidth of each resonance is
\eq{Dnu}{\delta\nu_{\frac{1}{2}}=\frac{c}{2LF}.} 
For a measurement of the difference between two resonance frequencies, the expected sensitivity is equal to the residual frequency instability of each laser multiplied by $\sqrt{2}$ (because a comparison is being made between two uncorrelated and equally noisy signals).  This gives a fractional frequency measurement sensitivity as:
\eq{dnurel}{\delta\nu_\mathrm{rel}&=&\sqrt{2}\delta_{\mathrm{shot}}\frac{\delta\nu_{\frac{1}{2}}}{\nu}\\
&\sim&\sqrt{\frac{h}{2\Pd \ti \nu}}\frac{c}{LF}.}
Using  experimentally realizable parameters, an indicative overall sensitivity can be given as:
\eq{Ndnurel}{\delta\nu_\mathrm{rel}&\approx&2.6\times 10^{-20} \,\,{\frac{3\unit{m}}{L}} \,\,{\frac{52\,000}{F}}\sqrt{\frac{5\unit{mW}}{\Pd}}\,\,\sqrt{\frac{1\unit{s}}{\ti}}}
In practice, to attain a shot-noise-limited measurement sensitivity it will be necessary to modulate the birefringence  at a judicious frequency that is well-removed from electrical or mechanical interference.  Although it is unlikely to expect shot-noise limited sensitivity at all frequencies it is certainly experimentally feasible to achieve this over a limited frequency band~\cite{day,bondu,sal,uehara, kawamura}.

\section{Measuring Optically-Induced Birefringence}
In order to measure a possible birefringence it is necessary to have an auxiliary `pump' laser beam to interact with the two pulsed detection beams in the RPI.  The detection beams do not act to  produce  birefringence upon themselves~\cite{partovi}.  It would be in principle possible to use a coaxial and counter-propagating pair of pump and detection beams and thus use a single set of mirrors for both the detection and pump processes.  However, it has been shown that this approach is potentially unsafe since dielectric mirrors can exhibit  a  strong photo-refractive effect, and this effect  can masquerade as a spurious birefringence signal by providing a means for the detection and pump beams to interact~\cite{hall}. 

We propose a second optical resonator to enhance the power of the pump beam, as illustrated in Figure~\ref{dualcavities}, which lies at an angle, $\theta$, with respect to the detection resonator axis. An additional advantage of this twin resonator approach is the ability to independently optimize the resonator mirror characteristics for the  detection and pump tasks.  For the calculations that follow the resonators are defined to be of identical length $L$, and we assume a separation between the resonator axes of $x$ at the cavity mirrors of radius, $a = x/2$ (see Fig.~\ref{dualcavities}). 
\makefig{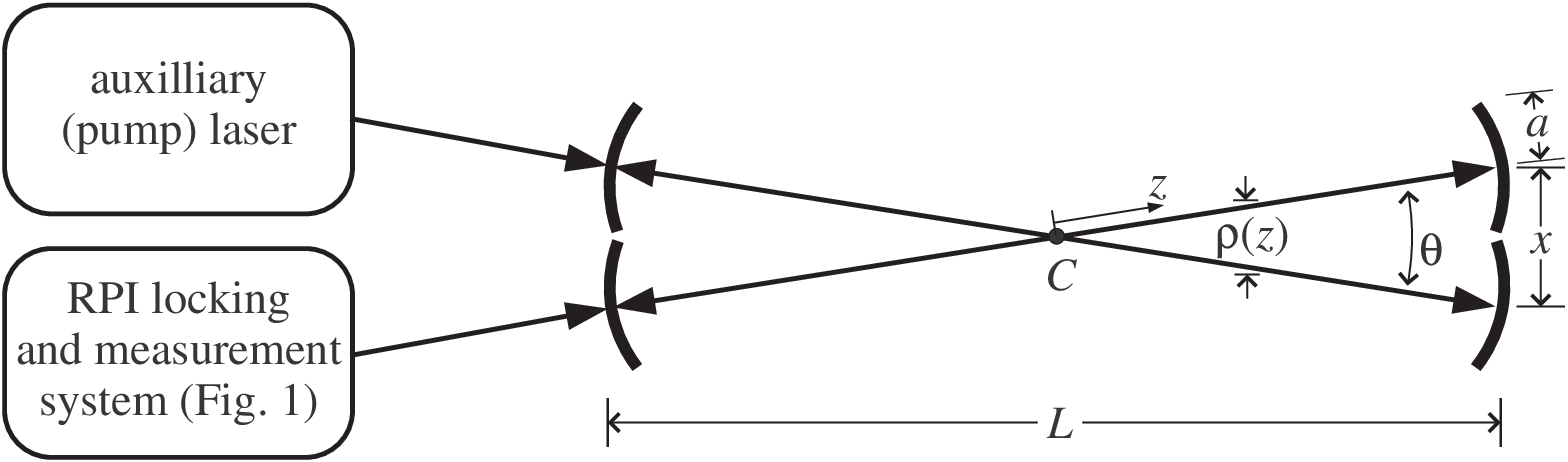}{dualcavities}{Measurement scheme for optically-induced birefringence.}  

Since the RPI produces a beat frequency corresponding  to the integrated birefringence in the cavity (see Eq.~(\ref{fracbirefringence})), a key concern is  the limited interaction region between the pump and detection beams. This length limitation   is  imposed by the crossed cavity design. 
It is one of the unique and key suggestions of this article that both the detection beams and the pump beam consist of laser pulse trains  rather than continuous-wave (cw) signals. If the timing of the circulating pulse in each beam is synchronized so that the detection and pump pulses meet head on at $C$ (see Fig.~\ref{dualcavities})~\cite{ma,holman,fortier,joneshol}, and in addition, each of the pulses is short enough to completely pass through each other before the beam axes begin to separate, then essentially all of the light circulating in the RPI cavity will interact with essentially all of the light circulating in the pump cavity on every pass.  Furthermore, the pulses pass through  each other  where the beams are most tightly focussed, and thus where they are most intense.  Although  the use of  pulsed lasers will complicate the experimental arrangement there is no ``in principle'' reason that a mode-locked laser signal can not be frequency locked with the same accuracy as a cw signal.  In fact,  mode-locked lasers have already been frequency-locked to resonators with relatively high precision~\cite{jones-lock,vidne}. A number of other authors have shown that low-dispersion resonators  can allow even very short pulses to be coupled into the resonator  with low power loss and relatively little broadening of the circulating pulse with respect to the input pulse~\cite{jones-dco,me-pulse}. An additional advantage of this pulsed-RPI approach is that we have automatically placed energy into many successive longitudinal modes of the detection and pump cavities. This circumvents a possible low frequency interaction between the  cw detection beams that has been seen in earlier  experiments~\cite{hall}.  The pulsed-RPI scheme automatically implements the more complex detection strategies proposed by Hall et al and Lee et al that avoids this issue~\cite{hall,lee2}.

To determine the potential sensitivity of the pulsed-RPI proposal we consider the case of a birefringence effect that is proportional to the intensity of the local optical field. This is true for both the Cotton-Mouton effect and the predicted QED vacuum polarization.  First it is necessary to determine the average intensity seen by a pulse circulating in the RPI cavity, which is equal to
		\eq{Iav}{\Iav&=&\frac{1}{L}\Lint{I(z)}}
where $I(z)$ is the intensity as a function of longitudinal position in the cavity.  When short pulses are used, it is merely  the region where the pulses pass through each other  that contributes significantly to the above integral.  This interaction region is approximately half the length of the pulses themselves, extending a distance 
		\eq{zodash}{z_0'  =   \frac{c \, \tau}{4}.}
either side of point $C$ on Fig.~\ref{dualcavities} where $\tau$ is the full width at half maximum pulse duration. As long as the separation, $\rho(z)$, between the beam axes remains significantly less than the beam radii in the interaction region, the beams can be treated as approximately coaxial when calculating \Iav.  The separation between the beam axes, in terms of beam radii, can be expressed as:
		\eq{rorel}{\frac{\rho(z)}{w(z)}&\apx&\frac{z \,  \theta}{w_0\sqrt{1+\brac{\frac{\lam z}{\pi w_0^2}}^2}}.}
where we are only interested in a range of $z$ that falls within the interaction region given by Eq.~(\ref{zodash}), $w_0$ is the beam waist size, and $\lambda$ is the wavelength of the stored radiation.
The minimum separation, $x$, between the mirror centers, as shown on Fig.~\ref{dualcavities}, is equal to twice the cavity mirror radius, $a$. The mirror radius must in turn be a factor of $\alpha$ larger than the laser mode spot radius evaluated at the  mirror location, $w(L/2)$,  where $\alpha$ is determined   by the extent to which aperture losses can be tolerated for a particular application. Thus $x$ is given by
		\eq{x1}{x=2a=2\,\alpha  \, w\brac{{L/2}}.}
Since we wish to maximize the induced birefringence we choose detection and pump cavity configurations that are close to the concentric stability limit~\cite{sieg},  so as to minimize the waist size in the cavity, and hence maximize the pump energy density.  In the limit of a small waist size, $w_0$, we can calculate the beam size at the mirrors and hence the required spacing between the mirror centers:
		\eq{x2}{x\approx \frac{\alpha L\lam}{\pi w_0},}
%
which  determines $\theta$, the angle between the beams as
		\eq{theta}{\theta=\frac{2\alpha\lambda}{\pi w_0}.}
For reasonable assumptions of a pulse duration below  200\,fs, a wavelength greater than  500\,nm, and a waist size greater than $5\lam$,  we find that the pulse length is less than one Rayleigh range, $z_\mathrm{RR} = \pi w_0^2/\lambda$. In this case we can simplify    
Eq.~(\ref{rorel}) and combine the result with Eqs.~(\ref{zodash}) and  (\ref{theta}) to give the following approximation for the relative separation:
	\eq{}{\frac{\rho(z)}{w(z)}\apx\frac{2 z\, \alpha\lam}{\pi w_0^2},}
	which becomes a maximum at the beginning and end of the interaction zone, $z = z_0'  =   c \, \tau/4$, 
		\eq{rorel2}{\frac{\rho(z_0')}{w(z_0')}\apx\frac{c\tau\alpha\lam}{2\pi w_0^2}.}
If the  relative separation is small at this point, then the  beams may be treated as coaxial over the entire region.  As an example, if $\rho(z_0')/w(z_0')=0.3$, then $I(z_0')$ is only about 6\% less on the detection axis than it is on the pump axis.  The reduction in average intensity when integrated over the entire interaction region is even smaller than this value.  To give a rough criterion for the minimum waist radius that can be used without encountering significant beam separation inside the interaction region, we set Eq.~(\ref{rorel2}) equal to 0.3 and rearrange, obtaining
		\eq{wo1}{
			w_0\gtrsim 10\micron
			\brac{\frac{\tau}{200\mathrm{\,fs}}}^{\frac{1}{2}}
			\brac{\frac{\alpha}{4}}^{\frac{1}{2}}
			\brac{\frac{\lam}{800\mathrm{\,nm}}}^{\frac{1}{2}}.}

We note that if the interaction region is smaller than the Rayleigh range of the beam, the beams will be of approximately constant radius as the pulses pass through each other.  A waist radius which is too small, though, will cause the beams to begin to diverge while still inside the interaction region and reduce \Iav.  Equating the Rayleigh range to $z_0'$ as given in (\ref{zodash}) and rearranging yields the following expression which must be satisfied in order to prevent significant beam divergence inside the interaction region.
		\eq{wo2}{w_0\gtrsim 2\micron\brac{\frac{\tau}{200\mathrm{fs}}}^{\frac{1}{2}}\brac{\frac{\lam}{800\mathrm{nm}}}^{\frac{1}{2}}.}
For realistic values of $\alpha$, adhererence to the inequality in Eq.~(\ref{wo1}) automatically satisfies (\ref{wo2}).

So long as the inequalities in Eq. (\ref{wo1}) and (\ref{wo2}) hold,  calculation of \Iav\ is straightforward.  Each time a detection pulse passes through the interaction region, it sees a burst of light which carries the effectively the entire energy \Ep\ of the pulse circulating in the pump cavity.   During the entire interaction time the pulses are approximately coaxial, with beam radii equal to that at  the waist.  Thus
\eq{Iav2}{\Iav&\apx&\recip{L}\frac{\log 2}{\pi w_0^2}\Lint{P(z)}\\
&\apx&\frac{c}{L}\frac{\log 2}{\pi w_0^2}\Ep.}
	
The circulating pulse energy, \Ep\, is determined by the average input power \Pav , the repetition rate, $R$, of the input pulse train, the resonator finesse $F$, and an efficiency factor $k_\mathrm{cav}$ which allows for mode-matching, impedance matching and dispersion related losses~\cite{jones-dco,me-pulse}: 
\eq{ep}{\Ep=k_\mathrm{cav}\,\frac{F}{\pi}\frac{\Pav}{R}.}
In addition, for the circulating pulse to be efficiently reinforced on each pass by the incident pulse train it is necessary that the free spectral range of the cavity is identical to the repetition rate of the laser, $R$~\cite{jones-dco,me-pulse,vidne}:
\eq{R}{R=\frac{c}{2L}.}
	In a time domain view this is equivalent to setting the inter-pulse time of the pulse train equal to the round trip time of the resonator.
Combining the three equations above gives:
\eq{Iav3}{\Iav\apx\frac{2\log 2}{(\pi w_0)^2}F\Pav.}
	
According to Eq.~(\ref{Iav3}), \Iav\ is determined solely by the pump oscillator average power, the pump resonator finesse and the size of the beam waist.  The beam waist, in turn, depends on $\lam$, $\alpha$ and $\tau$ via Eq.~(\ref{wo1}).  This results in the following indicative numerical expression for \Iav.
\eq{Iav4}{\Iav\apx \frac{F}{52\,000}\frac{\Pav}{20\unit{W}}\frac{200\unit{fs}}{\tau}\frac{4}{\alpha}\frac{k_\mathrm{cav}}{1}\frac{800\unit{nm}}{\lam}\times1.5\unit{\frac{PW}{m^2}}.}
	
The scaling factors chosen in Eq.~(\ref{Iav4}) reflect realistic experimental parameters.  A finesse of 52\,000 corresponds to a reflectance of $99.994\%$ which is available in a custom low dispersion mirror coating~\cite{ltech}.  These coatings have sufficiently low dispersion to allow 200\,fs incident laser pulses to be directly coupled into a cavity with near-unity efficiency~\cite{jones-dco,me-pulse}.  A mode-locked laser with a 200\,fs duration output pulse and  20\,W  average power has been reported with a repetition rate  of 25\,MHz~\cite{brunner}. It is likely that there will be further improvements in the output power of mode-locked lasers given the relatively early stage of development of this technology together with the rapidly decreasing cost of pump lasers.  Thus, using readily available equipment  it should be possible to construct a pump cavity which gives an effective average intensity in the detection cavity of  1.5\,$\mathrm{PW/m}^2$.  Such high average intensity is possible because by pulsing both the detection and pump beams, we have arranged for the detection pulses to see the same average applied intensity as if the beams were parallel and nondivergent throughout the cavity. It is  the pulsed and counter-propagating nature of the two beams that circumvents the effect of  high divergence which would normally undermine the use of tightly focussed light beams, and also ensures that the detection beam sees all the pump light on every round trip in the cavity. In fact, the pulsed beams show the same degree of  interaction as  cw beams that were parallel and nondivergent throughout the cavity, which is of course not possible for tightly focussed, non-coaxial beams.

The average intensity given by Eq.~(\ref{Iav3}) is independent of the length of the cavity, because the increase in energy per pulse that would occur if we switched to a lower repetition rate is cancelled by the decrease in the fractional length of the cavity which falls inside the interaction region.  We have assumed that the average output power of the pump laser is independent of the repetition rate, which is reasonably well followed by commercial laser systems. However, it should be noted that a longer cavity is preferable since it  increases the measurement sensitivity  (see Eq.~(\ref{Ndnurel})\,).  The optimal cavity length in a real experiment depends largely on the feasibility of constructing sufficiently large mirrors as implied by Eq.~(\ref{x2}).  For a 3 meter (50\,MHz repetition rate) cavity with $\alpha=4$, the mirrors would need to be approximately 20\,cm in diameter.  Although this presents a significant challenge, it is not insurmountable, as demonstrated by the recent construction of even larger diameter, high quality mirrors for gravitational wave detection interferometers~\cite{mirrors}. Alternatively, in order to reduce the size of the mirrors one can use more complex cavity geometries using two curved mirrors and two flat mirrors. 
	 
	%
	%
	One of the challenges of the concentric cavity required for this proposal is its sensitivity to misalignment of the cavity mirrors and pointing fluctuations of the input beam.  One can show that the waist size in a near concentric cavity is given by~\cite{sieg}:
	\eq{waistsize}{w_0 = \sqrt{\frac{R\,\lambda}{\,\pi}} 
  {\left( \frac{{\Delta L}}{2\,R - {\Delta L}}
      \right) }^{\frac{1}{4}}}
	where $\Delta L  = 2 R-L \ll R$, and $R$ is the radii of curvature of the two symmetric mirrors.  Thus in order to have a waist size of order 10\,$\mu$m in a cavity of length 3\,m it is necessary to tune the length to within 2~x~$10^{-7}$\,m of the instability limit.  In this near-concentric position the input coupling is highly sensitive to relative angular and translational misalignments of the cavity mode and the input beam mode.  The  beam displacement on mirror 1 or 2  given by~\cite{sieg}:
	\eq{displace}{{{{\Delta x}}_{(1,2)}} = 
  \pm \frac{R^2\,\left( -{{\theta }_{1}} + {{\theta }_{2}} \right) }
   {{\Delta L}}}
	where $\theta_{(1,2)}$ is the angular rotation of mirror 1 or 2.  In order to restrict translational motion of the mode on the mirrors to less than $1 \%$ of the spot size of the beam it is necessary to limit beam mode-cavity mode angular misalignments   to below 3~x~$10^{-10}$\,rad. 
 This alignment expression is clearly divergent as the cavity condition approaches the concentric limit. To meet the challenge of these alignment requirements one can either implement a mode-cleaning cavity before the detection and pump cavities~\cite{gossler}, or make use of an auto-alignment system~\cite{sampas}. We note that the requirements for mirror stability stated above are within the capability of such alignment systems~\cite{sampas}.
  
  We now turn our attention to higher order modes in this concentric cavity system.  The transverse mode spacing, $\Delta \nu_\mathrm{TM}$, in a near-concentric resonator can be found to be~\cite{sieg}: 
  	\eq{tms}{\Delta \nu_\mathrm{TM} = \frac{c}{4R} \left( 1 - {{}\,{\sqrt{\frac{{2\, \Delta L}}
         {R \pi^2 }}}} \right) }
        Using the example of a 3\,m near-concentric cavity  with a 10\,$\mu$m waist we find an $\sim$8\,kHz frequency splitting between the fundamental mode and the first-order transverse mode.  The higher order modes will be frequency resolved	if the finesse of the cavity is greater than 6,000.  In order that pointing fluctuations do not couple into frequency fluctuations it is necessary to have a finesse higher than this value.  In addition, we note that the transverse mode spacing is a useful diagnostic for setting the length of the cavity to achieve a desired cavity waist  size.
	
	%
	
	
\section{\label{sec4}Comparing pulsed-RPI to cw-RPI and  conventional Ellipsometry}
	

To place the sensitivity of the proposed pulsed RPI technique in context we should compare it with conventional resonant ellipsometry and with continuous-wave RPI (cw-RPI). In this section we also compare the energy density of an optical pump field  with that obtainable from a large scale static magnetic field.  
	
	Conventional resonant ellipsometry relies on `tuning' the rotational angle of the cavity mirrors  to set the intrinsic birefringence of the cavity to nearly zero for the linearly polarized input radiation~\cite{chui}. In this case  the slow or fast axis of each of the mirrors  is well aligned with the input polarization state  resulting in  limited conversion of the input radiation into   	the other polarization state by the mirrors.  
	The polarization of the input beam is set at $\pi / 4$
 with respect to the applied magnetic field direction using a high quality polarizer	and will become  elliptically polarized by the birefringence in the cavity.  The change in ellipticity of the beam can be expressed as:
 \eq{ellipticity}{\psi = k L \frac{\npar-\nperp}{2}.}
where $k$ is the wavenumber of the input light, and $L$ is the interaction length of the field and the light.  The induced ellipticity is measured by a polarization analyser at the output of the cavity. The analyzer consists of  a polarizer that has been set to pass radiation with polarization orthogonal to  the input radiation.   
The most sensitive measurements of ellipticity, $\psi$, using    single-pass ellipsometry (without any resonant cavity) have reported a $\psi$ detection limit of ~$10^{-8}/\sqrt{\ti}$, which is  less than a factor of 2 from the shot-noise imposed limit under their respective conditions~\cite{cameron,Muroo,muroo2}:

	 \eq{ellipNtoS}{\delta \psi_{\rm{ellips}} = \frac{\sqrt{2 +  4\, \sigma^2 / \theta^2}}{\sqrt{P\,
         \eta_{\rm{PD}}/ (h\,\nu)}} \frac{1}{\sqrt{\ti }}} 
	where  ${\eta }_{\mathrm{PD}}$ is the quantum efficiency of the detection photodiode and $\sigma^2$ is the extinction coefficient of the polarizer and analyzer. In order to linearize the sensitivity of the ellipsometer to small birefringence signals, and to shift the signal of interest away from zero frequency, it is usual to add a polarization modulation of depth  $\theta$ using a modulator at the output of the ellipsometer cavity (if present). In this case the birefringence signal now appears as sidebands about the modulation signal and can be demodulated using synchronous detection techniques.
	Unfortunately, the introduction of a resonant cavity or delay line into the ellipsometer in order to increase the length of the interaction between the applied field and detection beams ($L$ in Eq.~\ref{ellipticity}), and hence improve the birefringence sensitivity, results in a significantly worsened single pass phase  sensitivity~\cite{cameron}. The most sensitive birefringence measurements with delay lines or a high finesse cavity incorporated into the ellipsometer  have a birefringence sensitivity  in  the range of $\Delta n \sim 10^{-17} - 10^{-18}$~\cite{cameron,pvlas3}.	
	
	We now turn our attention to the RPI technique of measuring birefringence. We stated above in Eq.~(\ref{Ndnurel}) that an order of magnitude estimate of the shot-noise limited birefringence sensitivity was  below  $10^{-20}/\sqrt{\ti}$ where $\ti$ is the integration time of the measurement.  A more detailed examination of the sensitivity limits under Pound-Drever-Hall (PDH) locking to a cavity with perfect impedance and mode matching shows that~\cite{sal,black}:
	 \eq{PDHNtoS}{ \delta \psi_{\mathrm{PDH}} ={\frac{\sqrt{2}\,\pi \,{n_0}}
     {8\,{F}\,{\sqrt{\ti }}\,
       {\sqrt{{P\,{{\eta }_{\rm{PD}}}}
           /({h\,\nu })}}}}}
	
	To compare the sensitivity of the ellipsometer and RPI approaches we note that Eq.~(\ref{ellipNtoS}) represents the ellipsometry sensitivity for a single pass through the interaction zone whereas Eq.~(\ref{PDHNtoS}) naturally refers to a resonant measurement in a cavity of Finesse, $F$.  The sensitivity of a resonant   ellipsometer measurement  can be found by adjusting $L$ in Eq.~\ref{ellipticity} for the number of passes through the interaction zone, which for a resonator of finesse, $F$, will be a factor of  $2\,F/\pi $.  In the case where the intentional modulation depth in the ellipsometer is much greater than the extinction of the polarizer-analyzer pair ($\theta \gg \sigma$) then the sensitivity of the two approaches has an identical dependence on  the main experimental parameters with the RPI approach being $2 \sqrt{2}$ more sensitive. It is likely that  subtle technical details will be the ultimate determinant of which technique is optimal. 
	
	As an example of the types of experimental details which are of importance, the above expressions have excluded the effects of amplitude noise in the input laser beams. The two techniques will be sensitive to the amplitude noise in the immediate frequency environment of the modulation frequency. In the case of the ellipsometer this is    the  polarization modulation frequency, while  the phase modulation frequency inherent in a  PDH frequency lock  is the relevant parameter in the other case.  In a suitable resonator (where  $\delta \nu_{\frac{1}{2}} \gg \Delta \nu \frac{\delta \phi}{2 \pi}$) it is however possible  to have both systems deployed simultaneously~\cite{lee2}. 
 
 We note the analysis by Chui et al~\cite{chui} which compares the sensitivity of a continuous-wave(cw) RPI scheme and conventional resonant ellipsometer  to cavity mirror temperature changes.  In both schemes a mirror temperature change gives rise to a false birefringence signal although it is stated that the RPI approach is much more sensitive to these types of  temperature changes~\cite{chui}.  For the ellipsometer approach it is possible to reduce the sensitivity to temperature changes by $10^6$ times by accurate alignment of the input beam direction with the intrinsic birefringence axis of the mirror surfaces.  In the worst case the cw-RPI technique will require  the intrinsic birefringence of the mirrors to be stable to 1 part in $10^{11}$ during the measurement period, which  corresponds to a temperature stability for the mirrors in the 10$^{-9}$\,K range.    Although this appears to be an extreme challenge for the RPI approach we point out two  important differences in our scheme in comparison to that considered by those authors. First, it is possible to choose birefringence matched mirrors and align the slow axis of one mirror with the fast axis of the other mirror in construction of the detection cavity.  In this case the frequency difference between the two polarisation modes of the cavity will be much reduced, which reduces the temperature stability requirements by the same large factor (if the temperature fluctuations of the mirrors are correlated). In addition, as will be pointed out below, in the case of a pulsed RPI system it is possible to modulate the effective `pump' intensity at a high frequency ($>10$\,kHz) (unlike the assumption of   Chui et al. that has modulation frequencies of $\sim$1\,mHz). Slow temperature changes of the mirrors will be very strongly suppressed by this modulated measurement technique.  
	
	We now turn our attention to the magnitude of the  polarizing field (the `pump' beam).
	An average detected intensity of 1.5\,$\mathrm{PW/m}^2$ (see Eq.~(\ref{Iav4})\,) corresponds to an energy density of 5\,$\mathrm{MJ/m}^3$, a little lower than the 39\,$\mathrm{MJ/m}^3$ produced by a 10\,T laboratory magnetic field. Thus focussed short pulses of light  are only slightly lower in energy density than the conventional magnetic field approach.  
	 We note in passing that extremely high intensity fields (much higher than can be generated by any macroscopic magnetic field technique) can be created by tightly focussing the output of  a high energy laser pulse amplifier~\cite{lee}.   The difficulty with this approach lies in constructing a detection system with sufficient sensitivity to probe inside these short pulses given the tight temporal and spatial restrictions~\cite{lee}.  In addition, these high pulse energy amplifiers have relatively low repetition rates limiting the measurement rate.
	
	Finally, we comment that one of the very great advantages of pulsed-RPI over cw-RPI is the ability to modulate the effective strength of the pump field at a high and almost arbitrary rate without varying the energy load or distribution on the mirror surfaces.  This enables detection of the birefringence signal in a frequency domain where there is minimal noise interference, without giving rise to potentially false signals. We achieve this effective power modulation by temporally delaying or advancing the pump pulse with respect to the detection pulse and thus varying the degree of energy overlap at the crossing point of the two cavities.  This type of power modulation results in no change on the thermal load of the mirrors and thus eliminates many potential spurious effects that could otherwise masquerade as the effect of interest.  This technique can be implemented as part of the control system that synchronizes the detection and pump pulses~\cite{ma, holman,joneshol}.

\section{\label{sec5}Detecting Vacuum Birefringence}
		
A birefringence effect of significant interest at this time is that arising from a scattering of photons from a static electric or magnetic field, or even from other real photons. Although it was predicted almost seventy years ago that virtual positron-electron pairs in the quantum electrodynamic vacuum could mediate interactions between photons~\cite{euler, heisenberg, weisskopf,schwinger}, this effect has yet   to be observed directly in the laboratory as a refractance or birefringence of the vacuum.  Nonetheless, there is evidence of scattering of photons from extremely strong electric fields and inelastic photon-photon scattering  in high-energy physics experiments~\cite{rohrlich,wilson,burke}.
It is believed that vacuum polarization plays an important role in extreme astrophysical environments such as exist at the surfaces of pulsars~\cite{heyl}. 

The QED-mediated interaction between a polarized field and a polarized photon gives rise to a polarization-dependent optical refractance of the vacuum.  For the effect to be induced by an optical field, a linearly polarized `pump' beam must interact with a counter-propagating `detection' beam.  The detection beam can then be regarded as moving in the mean field  of the pump beam.  The refractive indices of the vacuum for light polarized parallel and perpendicular to the polarization of the pump beam are denoted as \npar\ and \nperp, and are given by \cite{alek}
\eq{nn}{\npar=1+\frac{16}{45}\frac{\alpha^2U}{U_e};\;\;\;\;\;\nperp=1+\frac{28}{45}\frac{\alpha^2U}{U_e}}
where $\alpha$ is the fine structure constant, $U$ is the energy density in the optical field and
 ${U_e={m_e^4 c^5}/{\hbar^3}\approx 1.42\times 10^{24}}$\,J/m$^3$
is the Compton energy density of the electron ($m_e$ is the electron rest mass).  Equation~(\ref{nn}) demonstrates that the induced refraction is polarization dependent and hence the vacuum exhibits both  a change in the phase velocity of the detection light because of the presence of the pump beam but also a  birefringence given by
\eq{b}{\Delta n=\frac{4}{15}\frac{\alpha^2\,U}{U_e}=\frac{4}{15}\frac{\alpha^2\Iav}{c\,U_e}.}
Although these expressions only strictly hold for infinite plane waves, they give a birefringence of the correct order of magnitude so long as the beams remain well-collimated over the interaction region.  Substituting the maximum average intensity which can be applied in the detection cavity of the pulsed RPI from Eq.~(\ref{Iav4}) into Eq.~(\ref{b}) gives an estimate of the expected  birefringence.

Various challenging technical issues must be addressed in order to implement this experiment although we note that many of the elements of this experiment have been demonstrated elsewhere. For example, the pulse trains must be appropriately synchronized so that the pulses meet where the beam axes cross~\cite{ma,holman,joneshol}.  In addition, the offset frequency and repetition rate of the outputs of the pulsed lasers must be controlled to match the cavity resonance frequencies and free spectral range of both cavities~\cite{jones-lock}, while both the detection and pump cavities must have the same free spectral range.  The final hurdle will be the duration of the experiment observation time in order to unambiguously detect  the effect.  We calculate these integration times  by equating the expression for shot-noise limited measurement sensitivity in Eq.~(\ref{PDHNtoS}) with the expected vacuum birefringence signal in Eq.~(\ref{b}) and present them in Table~\ref{tab1}.  The first two lines predict the performance available from existing low dispersion mirrors. The first line of the table show the performance capability of the best ``off-the-shelf'' commercially available low dispersion mirrors while the second line shows the capability of the best custom built mirrors.  Resonators built from   these mirrors are capable of accepting 200\,fs pulses without significant temporal distortion~\cite{jones-dco,me-pulse}. The last line of the table predicts that performance that would be available if  low dispersion mirrors would have a reflectivity equal to that of   the best commercially-available super-mirrors. 


\begin{table}[tb]
\centering
\begin{tabular}{|c|c|c|}
\hline
$R$, \%&$F$&$\tau_{int}$\\
\hline
99.97&1.0\ee{4}&2.6 years\\
99.994&5.2\ee{4}&1.7 days\\
99.997&1.0\ee{5}&2.5 hours\\
\hline

\end{tabular}
\caption{Required integration time \ti, for the detection of vacuum birefringence assuming a 20\,W, 200\,fs pump laser launched into a 3\,m long resonator tabulated as a function of the resonator mirror  reflectivity $R$ (it is assumed the mirrors of the pump and detection cavities are identical).}
\label{tab1}
\end{table}
  
 The measurement time required to detect \vb\ scales with the inverse fourth power of the finesse because the finesse affects both the measurement sensitivity and the average intensity in the pulsed RPI approach.  Competing techniques that rely on a macroscopic magnetic field to create a vacuum polarization have  an integration period that decreases only as the square of the finesse of the detection cavity. Thus improvements in mirror technology will result in the pulsed RPI technique soon outpacing competing strategies.  If low dispersion mirrors could be improved to the point that 99.997\% reflectivity mirrors become available (as good as existing super-mirrors) then the corresponding increase in finesse would allow vacuum birefringence to be detected in just a few hours.  This analysis neglects the likely increases in available laser power over the next few years which will also reduce the required measurement time.
 
 Despite the difficulties that could be expected in operating an optical system based on near-concentric cavities of such a large size, an all-optical device should be smaller, cheaper, easier to operate and more reliable than systems using helium-cooled superconducting magnets.  In addition, there are a couple of extremely important benefits accruing from the use of an optical pump field. First, there is the possibility to modulate the effective strength of the pump field at high rates as mentioned above, without changing the thermal load on the mirror system. Second,  the low forces and power required to generate   high intensity pulsed optical fields, combined with the high confinement potential of optical fields enables the elimination of many effects in the detection system which masquerade as a \vb\ signal in contemporary experiments~\cite{cameron, itnoise}. Finally, we would suggest that there will be a rapid development of optical and laser technology over the next few years, especially in low dispersion mirrors with high reflectance and the development of higher average power  mode-locked lasers. These developments will directly feed into an improvement in the performance of \vb\ detection system based on these types of technology.  We would not expect the same rate of development in superconducting electromagnet  technology.

\section{Conclusion}

We have proposed a new approach to the experimental detection of very low levels of field-induced birefringence.  In particular, we analyze the system for its applicability to direct detection  of the predicted vacuum nonlinearity. We believe that this system offers a strong possibility of being the first to detect this  effect. Our approach is based on the intersection of two concentric and high finesse short-pulse resonant cavities, one of which pumps the vacuum to produce the birefringence, while the second detects this induced birefringence using highly sensitive frequency metrology techniques. We predict a sensitivity that will  allow an experimental detection of the predicted vacuum nonlinearity after a measurement period of just a few days. This is a comparable  period to that predicted for conventional techniques, however, this new approach avoids the masking effects of spurious signals that plague conventional experiments.  A successful detection of this effect will enable a sensitive experimental test of a major prediction of Quantum Electrodynamics.
 
 \section{Acknowledgements}
 
 We thank the Australian Research Council (ARC) for financial support of this research.  We would like to thank all the members of the Frequency Standards and Metrology Group  which make it such a pleasant and stimulating environment in which to work.  In particular, we would like to thank John Winterflood, John McFerran and Sam Dawkins for reading the manuscript and providing useful feedback to the authors.

\end{document}